\documentclass[pdflatex,sn-mathphys-num]{sn-jnl}
\usepackage{bm}
\usepackage{graphicx}%
\usepackage{multirow}%
\usepackage{amsmath,amssymb,amsfonts}%
\usepackage{amsthm}%
\usepackage{mathrsfs}%
\usepackage{float} 
\usepackage[title]{appendix}%
\usepackage{xcolor}%
\usepackage{textcomp}%
\usepackage{manyfoot}%
\usepackage{placeins} 
\usepackage{booktabs}%
\usepackage{algorithm}%
\usepackage{algorithmicx}%
\usepackage{algpseudocode}%
\usepackage{listings}%
\usepackage{flafter}%
\usepackage{placeins}%
\newcommand{\emsp}{\hspace{1em}}

\theoremstyle{thmstyleone}%
\theoremstyle{thmstyletwo}%

\theoremstyle{thmstylethree}%

\begin{document}

\title[Article Title]{LLM-CoOpt: A Co-Design and Optimization Framework for Efficient LLM Inference on Heterogeneous Platforms}


\author [1]{\fnm{Jie} \sur{Kong}}\email{jiekong0112@sdust.edu.cn}

\author [1]{\fnm{Wei} \sur{Wang}}\email{wei\_wang@sdust.edu.cn}

\author [1]{\fnm{Jiehan} \sur{Zhou}\textsuperscript{*}}\email{jiehan.zhou@sdust.edu.cn}

\author [2]{\fnm{Chen} \sur{Yu}}\email{yuchen@hust.edu.cn}

\affil {\orgdiv{School of Computer Science and Engineering}, \orgname{Shandong University of Science and Technology}, \city{Qingdao}, \country{China}}

\affil {\orgdiv{School of Computer Science and Engineering}, \orgname{Huazhong University of Science and Technology}, \city{Wuhan}, \country{China}}


\abstract{Major challenges in LLMs inference remain frequent memory bandwidth bottlenecks, computational redundancy, and inefficiencies in long-sequence processing. To address these issues, we propose LLM-CoOpt, a comprehensive algorithm-hardware co-design framework aimed at improving both throughput and latency in LLM inference. LLM-CoOpt integrates three key strategies: (1) Key-Value Cache Optimization, termed Opt-KV, which improves memory access efficiency by optimizing both KV cache write and read paths, and introduces FP8 quantization to reduce memory footprint while maintaining accuracy; (2) Grouped-Query Attention for Computational Efficiency, termed Opt-GQA, which reduces the overall computational complexity by restructuring multi-head self-attention into grouped-query attention with shared key-value projections, enabling higher throughput and lower resource consumption; (3) Paged Attention for Long-Sequence Processing, termed Opt-Pa, which adopts a two-step strategy to first segment long sequences into manageable chunks and then apply lazy memory mapping and computation, significantly reducing memory pressure and improving performance on long-context inputs. Experiments on the LLaMa-13B-GPTQ model demonstrate that LLM-CoOpt increases inference throughput by up to 13.43\%, reduces latency by up to 16.79\%, and maintains model accuracy. These results confirm that LLM-CoOpt provides a practical, high-performance optimization path for real-world inference of large-scale language models.}




\maketitle

\section{Introduction}\label{sec1}

In recent years, Large language models based on the Transformer architecture \cite{1}, \cite{2}, such as LLaMa \cite{3}, ChatGPT \cite{4}, PaLM \cite{PaLM}, BLOOM \cite{BLOOM}, and BERT \cite{5}, have made breakthroughs in the field of natural language processing. However, the exponential growth of their scale has brought significant inference challenges, particularly on resource-constrained hardware. These challenges can be summarized into three major issues:\par

\begin{figure}[H]
	\centering
	\includegraphics[width=0.8\linewidth]{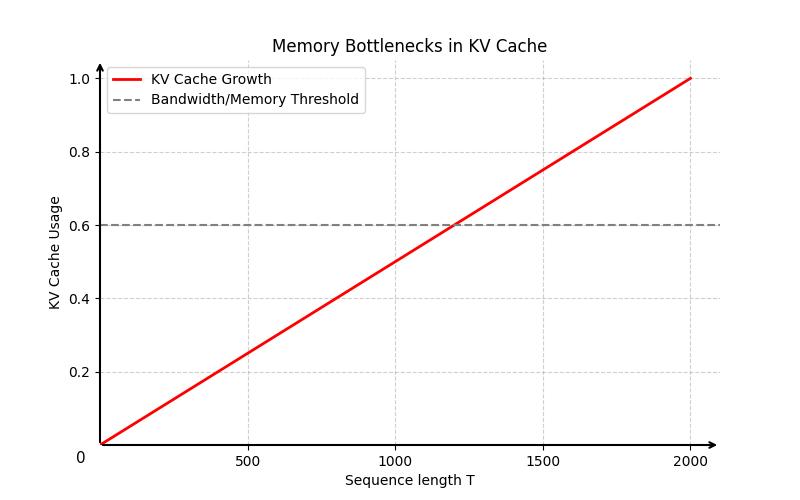}
	\caption{Memory Bottlenecks in KV Cache}
	\label{Fig1}
\end{figure}

\textbf{(1) Memory bottlenecks in KV Cache.} 
During auto-regressive decoding, the Key-Value (KV) cache grows linearly with sequence length, number of layers, and hidden dimensions. This not only leads to memory bloat but also imposes heavy pressure on bandwidth, since every new query must access all historical keys and values. On heterogeneous platforms, such inefficiencies in memory utilization directly degrade throughput, as illustrated in Fig.~\ref{Fig1}

\begin{figure}[H]
	\centering
	\includegraphics[width=0.8\linewidth]{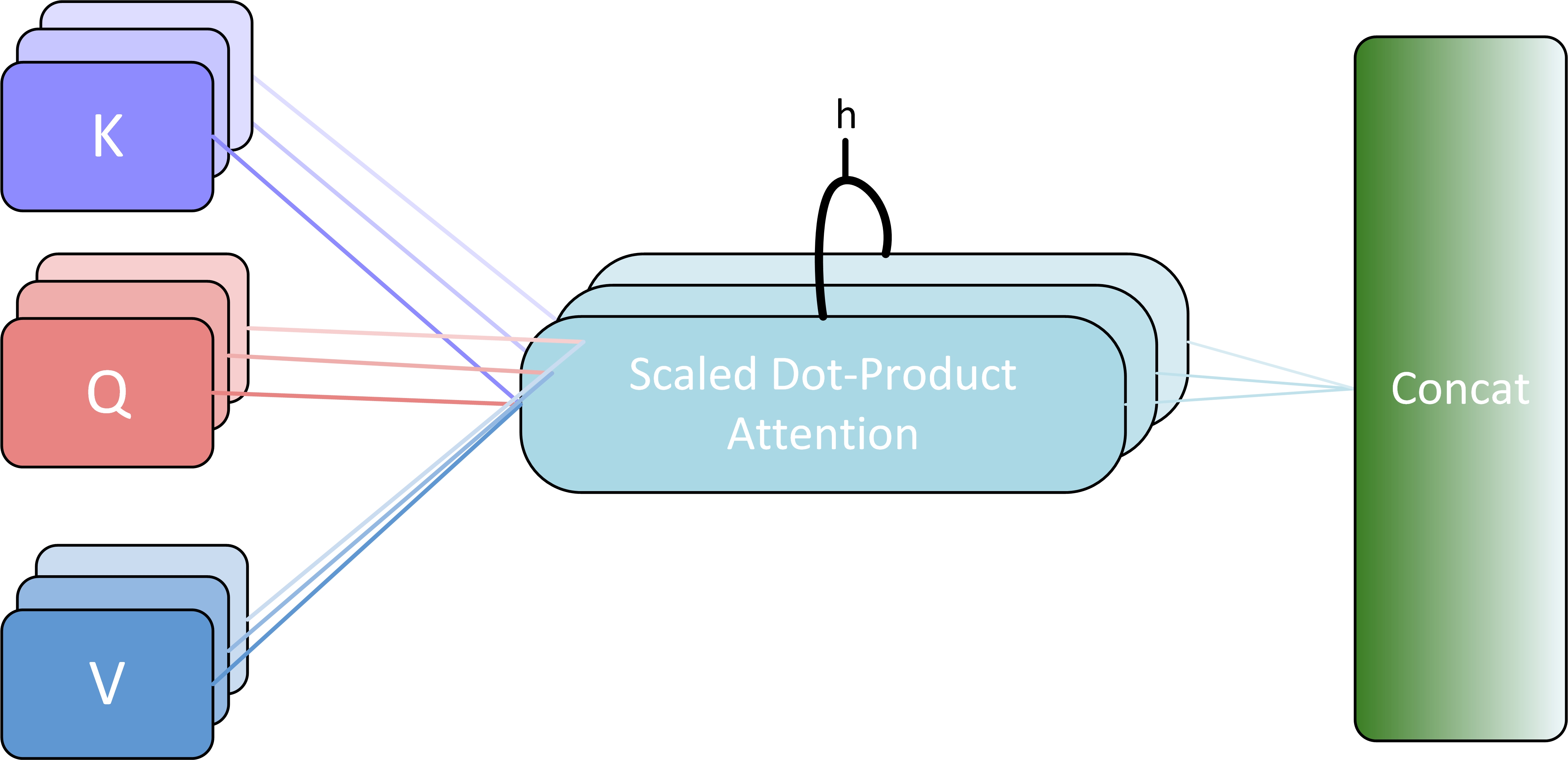}
	\caption{MHA: multi-head attention(Based on~\cite{2}).}
	\label{Fig2}
\end{figure}

\textbf{(2) Computational redundancy in multi-head attention.} 
In the standard Transformer, each attention head independently generates key–value pairs, leading to redundant computation and low utilization of parallel hardware resources. Although methods such as Grouped Query Attention (GQA) partially alleviate this problem, their fixed grouping strategies fail to adapt to different sequence distributions and hardware parallelism, leaving significant inefficiency in practice, as illustrated in Fig.~\ref{Fig2}

\begin{figure}[H]
	\centering
	\includegraphics[width=0.8\linewidth]{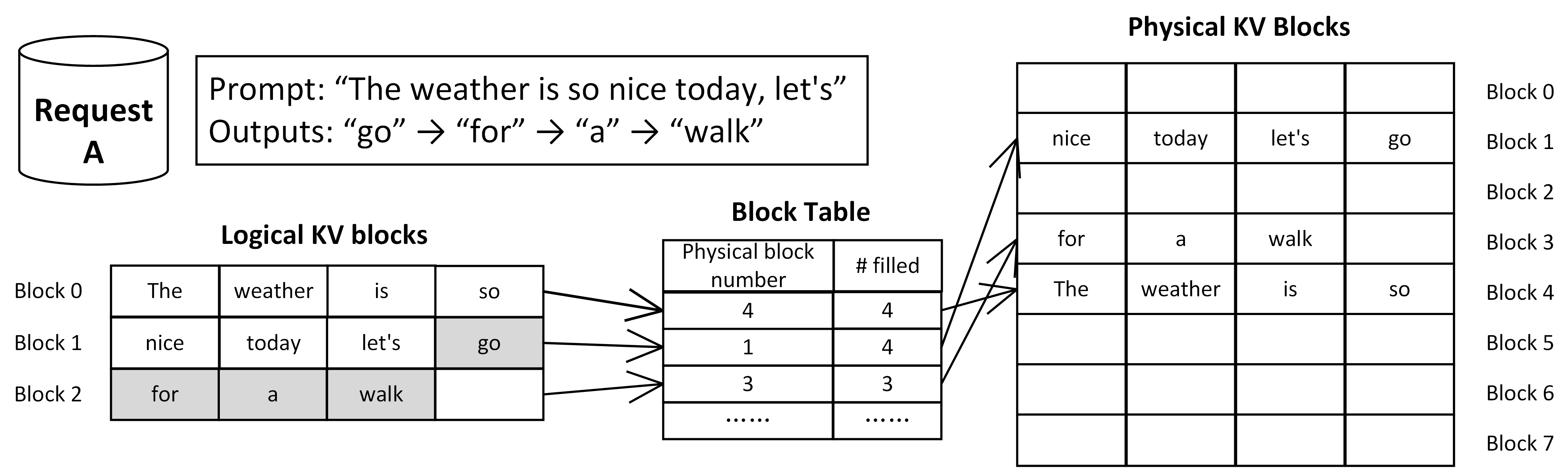}
	\caption{Storage Fragmentation Issue}
	\label{Fig3}
\end{figure}

\textbf{(3) Inefficiency in long-sequence processing.}
When processing long sequences, self-attention suffers from quadratic complexity ($O(T^2)$) and system inefficiencies. Paged attention, though reducing memory growth, adds synchronization overhead and fragmentation, causing unstable memory use and fluctuating latency. As shown in Fig.~\ref{Fig3}, such issues make long-context inference inefficient and unstable, limiting real-time deployment.
\par

Despite considerable research efforts, existing solutions have only partially addressed these challenges, often introducing new trade-offs in accuracy, efficiency, or hardware adaptability. Current studies can be broadly categorized into three directions: KV cache compression, redundancy reduction in multi-head attention, and efficient mechanisms for long-sequence modeling.\par

For KV cache management, static quantization has been applied to reduce memory usage \cite{6}, but it struggles to adapt to the varying dynamic ranges of different tensors \cite{44,45}. For instance, the distributions of query vectors and attention weights differ significantly, and uniform quantization exacerbates the trade-off between accuracy and efficiency \cite{39}. The FP8 format \cite{7} alleviates storage overhead but fails to meet the precision requirements of Softmax in attention computation, leading to numerical instability. Tao et al. \cite{13} further explored asymmetric quantization to enable 1-bit compression while preserving performance.\par

In terms of computational redundancy, the standard multi-head attention mechanism incurs significant overhead since each head independently generates key–value pairs \cite{8,40}. Grouped Query Attention (GQA) \cite{9,14,15} reduces this redundancy by sharing key–value heads, yet its fixed grouping strategy cannot adapt to dynamic input characteristics or hardware parallelism, resulting in suboptimal resource utilization.\par

For long-sequence modeling, chunked or localized attention \cite{10,46} reduces memory usage but compromises contextual coherence due to missing cross-block dependencies. Unified sparse attention \cite{16} and paged attention frameworks \cite{17,18,19} improve scalability but still face synchronization overhead, fragmentation, or accuracy degradation. More recent efforts, such as JAQ \cite{11} and hybrid SPM-Cache \cite{12}, attempt to balance efficiency and hardware friendliness, but at the cost of increased system complexity. Overall, while these approaches provide partial relief, none fully resolve the inherent conflicts between accuracy, efficiency, and hardware adaptability \cite{42}.\par

Based on the insights above, we propose LLM-CoOpt, an algorithm–hardware co-optimization framework that balances inference efficiency and accuracy for large-scale language model inference. The framework introduces three key techniques: Opt-KV for dynamic quantization cache optimization, Opt-GQA for lightweight grouped query attention, and Opt-Pa for paged attention optimization. Together, these innovations alleviate bottlenecks in memory bandwidth, computational redundancy, and long-sequence processing. Experiments on the LLaMa family show that LLM-CoOpt achieves significant throughput gains and latency reduction while maintaining strong accuracy on the ARC evaluation set.
\par

The remainder of this paper is organized as follows: Section 2 reveals performance bottlenecks when migrating Very Large Language Model (vLLM) to heterogeneous platform. Section 3 introduces the LLM-CoOpt framework and details. Section 4 describes the experimental setup and the evaluation of the results. Finally, Section 5 concludes the paper.

\section{Background}\label{sec2}

vLLM is a high-performance inference engine specifically optimized for large-scale language model inference workloads, characterized by high throughput, low latency, and excellent scalability \cite{37,38}. By incorporating the PagedAttention mechanism \cite{20}, vLLM enables dynamic and efficient management of the KV cache~\cite{43}. It also supports a range of optimization strategies such as dynamic batching and sequence merging \cite{21}, which collectively alleviate memory fragmentation and reduce cache redundancy, thereby significantly enhancing hardware resource utilization. When deployed on a single GPU, vLLM demonstrates $2\times{-}3\times$ higher inference throughput and lower response latency compared to the HuggingFace Transformers framework. Moreover, vLLM is compatible with low-bit quantized models, such as those using INT4 \cite{22,23}, enabling higher inference efficiency with negligible accuracy degradation \cite{35,36}, making it particularly well-suited for deploying large-scale models in resource-constrained environments.\par

However, when migrating vLLM from traditional GPU platforms to the heterogeneous platform, multiple system-level bottlenecks and adaptation challenges emerge. Although the heterogeneous platform offers strong computational capabilities and supports low-precision arithmetic, its limited memory resources and significantly different memory hierarchy compared to mainstream GPUs lead to a series of performance bottlenecks during the migration process. To further clarify the root causes of these bottlenecks, the following sections analyze the system limitations exposed on heterogeneous platform by examining each stage of the vLLM inference pipeline.\par

In the traditional vLLM inference pipeline, the inference process starts with token generation. During the token insertion phase, vLLM handles the KV cache using a block-based memory management strategy, where the KV tensor of each newly generated token is written to a newly allocated memory block. On the heterogeneous platform, direct migration to heterogeneous platform suffers from allocator inefficiency and increased latency due to allocator mismatch.\par
Secondly, during the attention computation phase, all historical KV blocks are uniformly loaded into explicit memory without priority or differentiation, and then matrix multiplication operations are performed as shown in Eq.~1, which will result in all KVs being loaded into memory regardless of whether they are actually useful or not, including padding and duplicate tokens, which will result in saturated memory bandwidth. Also, the heterogeneous platform's current memory is very limited compared to that of GPUs, making it prone to cache invalidation or overflow. Moreover, the computational workload increases linearly with sequence length, exhibiting complexity of $O(t \times d)$. Finally, K/V tensors are stored in FP16 or FP32 formats by default, which results in large data sizes and high memory usage that cannot efficiently adapt to the low-bit arithmetic supported by heterogeneous platform.

\par

\begin{equation}
	\mathrm{Attn}\left(q_t, K_{\leq t}, V_{\leq t}\right) = 
	\mathrm{Softmax}\left( \frac{q_t \bm{K}^\intercal}{\sqrt{d}} \right) \bm{V}
\end{equation}

Where $q_t$ is the query vector, $K_{\le t}$ is the key matrix, $V_{\le t}$ is the value matrix, and d is the dimensionality of the vector.\par
Compared to the traditional cache allocation strategy, the entire history block's data is multiplied, and even parts of the block that do not affect the attention score are indiscriminately allocated to the cache. This leads to many unnecessary computations being executed by the Single Instruction Multiple Data (SIMD) units inside the heterogeneous platform, reducing overall arithmetic utilization. Furthermore, it complicates the introduction of a dynamic cache management strategy. Additionally, invalid and valid tokens are mixed within the same logical page, increasing the overhead of subsequent data transfer and conversion. As a result, the computational workload grows linearly, as shown in Eq.~2.

\begin{equation}
	Used\ Cache=R\times S_{block}
\end{equation}

Finally, when reading tokens, the system adopts a hierarchical storage structure consisting of L1, L2 caches and Synchronous Dynamic Random Access Memory (DRAM), with different capacity and access latency characteristics of each cache layer, in which the L1 and L2 caches are fast to access but have limited capacity, while the DRAM has a large capacity and high access latency. The average load latency is shown in Eq.~3, $T_{DRAM}$ is around 400 cycles, there is the problem of low cache hit rate or critical metadata is not preloaded, and the actual latency will be close to the access latency of DRAM, which makes the kv data loading become a bottleneck. At the same time all blocks are moved into the kernel, with the load equation shown in Eq.~4.

\begin{equation}
	T_{effective}=H\times T_{Cache}+(1-H)\times T_{DRAM}
\end{equation}

Where H is the cache hit rate, $T_{Cache}$ is the cache access latency, and $T_{DRAM}$ is the DRAM access latency.

\begin{equation}
	C_{kernel}=B\bullet N_{block}\bullet d^2
\end{equation}

Where B represents the batch size, $N_{block}$ represents the number of blocks assigned to each sequence, and d represents the vector dimension of each attention header.\par

To alleviate the system bottlenecks revealed in the stages above and further improve the adaptability and execution performance of vLLM on the heterogeneous platform, it is essential to focus on the performance-critical paths in the inference pipeline. Performance profiling reveals that certain modules are invoked frequently and consume significant resources during inference, thus becoming the "hot functions" that most impact overall performance.\par

Recent studies have introduced targeted optimizations of the KV cache, PagedAttention, and GQA modules have been introduced to address key system-level challenges, including memory bandwidth bottlenecks, co-scheduling of compute and memory resources, management of large-scale datasets, and inefficient memory access patterns. These optimizations effectively reduce resource wastage and computational redundancy, resulting in a substantial improvement in the overall efficiency of the inference framework.

\section{Method}\label{sec3}

To cope with the problem of inefficient attention computation and memory access, we propose the co-optimization framework LLM-CoOpt, which is designed for reasoning on large language models on hardware-constrained platforms. The framework consists of three approaches: Opt-KV optimizes KV cache accesses to improve memory bandwidth utilization; Opt-GQA employs grouped query attention to enhance head flexibility and reduce computation overhead; and Opt-Pa improves paging strategy to enhance memory paging efficiency and data locality.

\subsection{Opt-KV}

The Opt-KV method enhances memory efficiency and reduces computation in attention inference via two stages: selectively caching only useful KV blocks while skipping redundant ones to cut memory writes, then compressing valid blocks into FP8 format to save storage. It also supports on-demand FP8 decompression during inference, ensuring efficient memory access without losing accuracy.
\par

To clarify Opt-KV’s optimization, we formally describe its low-bit memory-efficient read and write mechanism. It involves two parts: skipping redundant KV blocks during writes, and enabling online dequantization during reads. In the write phase, for token index i, the method checks a condition, as shown in Eq.~5; if satisfied, the corresponding KV block is skipped.

\begin{equation}
	\mathit{slot\_idx\_i} < 0 \ \mathrm{or}\ \mathit{slot\_idx\_i} \in \mathrm{SkipSet}
\end{equation}

In the read phase, cached key/value (KV) tensors are dequantized from FP8 format before attention computation. The restored key ${\widetilde{k}}_i$ and value ${\widetilde{v}}_i$ are computed as shown in Eq.~6.

\begin{equation}
	{\widetilde{k}}_i=\mathrm{dequant}(k_i^{\mathrm{FP8}}),\emsp{\widetilde{v}}_i=\mathrm{dequant}(v_i^{\mathrm{FP8}})
\end{equation}

This process is realized by a dedicated kernel function gather\_cached\_kv, ensuring data consistency and numerical correctness.

\begin{algorithm}[htbp]
	\caption{Opt-KV: KV Cache Optimization}
	\label{alg:opt-kv}
	\begin{algorithmic}[0]  
		\State \textbf{Input:} $Q$, $K$, $V$, $H$, $t$, $B$
		\State \textbf{Output:} $O$
		\Statex
		\State \textbf{Phase 1: KV Cache Write Optimization}
		\For{each token $i$}
		\If{$\text{slot\_idx}_{i}$ satisfies filtering condition}
		\State skip caching of $K_{i}$, $V_{i}$
		\Else
		\State reshape and store $K_{i}$, $V_{i}$ into block\_wise cache structure
		\EndIf
		\EndFor
		
		\State \textbf{Phase 2: FP8-Aware KV Retrieval}
		\For{each token $i$}
		\State Load cached $K_{i}$, $V_{i}$ based on mapped slot
		\State cache uses FP8
		\State Apply on-the-fly dequantization
		\EndFor
	\end{algorithmic}
\end{algorithm}

\subsection{Opt-GQA}

\begin{figure}[H]
	\centering
	\includegraphics[width=0.8\linewidth]{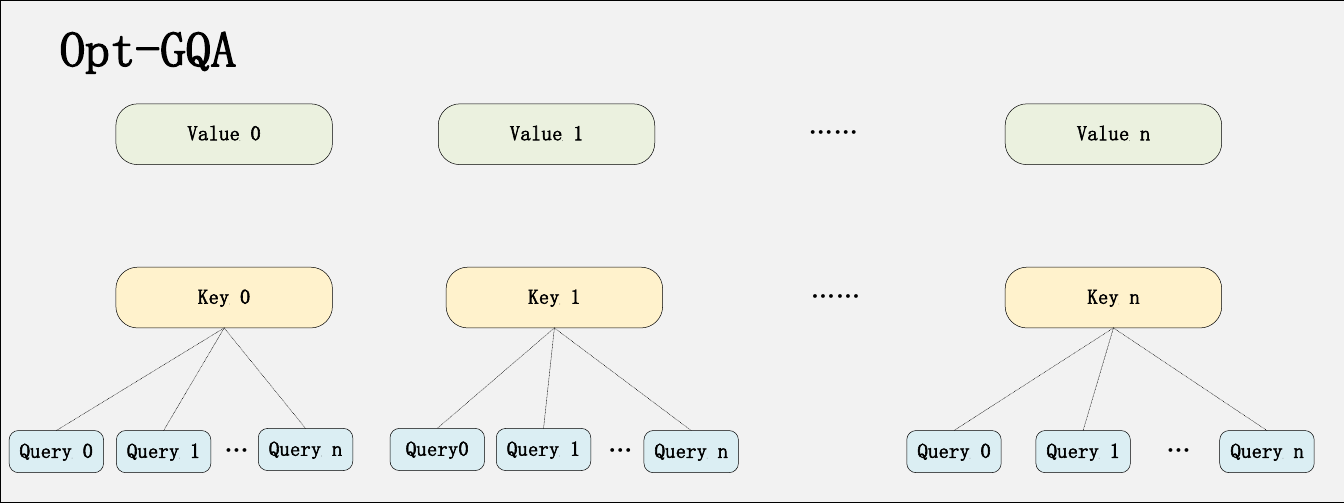}
	\caption{Opt-GQA: A Lightweight Grouped-Query Attention Design}
	\label{Fig4}
\end{figure}	

Fig.~\ref{Fig4} illustrates the Opt-GQA method, which enhances attention flexibility and memory efficiency by enabling grouped-query attention. Instead of computing attention over all heads uniformly, the method divides the query heads into multiple groups. Each group shares a corresponding key-value head, thereby reducing memory duplication and aligning the attention computation with the grouped query structure.\par

The diagram consists of several key components: Each block at the top represents a query group, while the middle layer shows the associated key-value heads. Each key-value pair is shared among all query heads in its group. The bottom layer illustrates the distribution of query heads mapped to respective key groups. This grouped computation structure provides a trade-off between flexibility and efficiency in attention modeling.\par
To provide a clear mathematical formulation of Opt-GQA, we describe its grouped-query attention mechanism, which improves attention flexibility and efficiency by partitioning query heads into groups that share the same KV heads. The query group mapping is defined as shown in Eq.~7.

\begin{equation}
Group_q(i)=\left\lfloor\frac{i}{H_g}\right\rfloor,\emsp\mathrm{where\ }H_g=\frac{H_q}{H_k}
\end{equation}

Where, $H_q$ denotes the number of query heads, and $H_k$ is the number of KV heads. $H_g$ thus represents the number of query heads per group. The group index for the $i-th$ query head is determined accordingly.

The attention weight for each query is computed using grouped keys, as shown in Eq.~8.

\begin{equation}
	\alpha_i = \frac{\exp\left(q_i \cdot k_{g(i)} - \max_{j \in g(i)} (q_i \cdot k_j)\right)}{\sum_{j \in g(i)} \exp\left(q_i \cdot k_j - \max_{j \in g(i)} (q_i \cdot k_j)\right)}
\end{equation}

Where $q_i$ is the query vector at position i, and $k_{\left\{g\left(i\right)\right\}}$ is the corresponding key from its group. The max$(q_i\cdot k_j)$term ensures numerical stability during softmax normalization.

Compared with traditional multi-head attention where each query uses its own KV pair, Opt-GQA enables multiple queries to share the same KV group, effectively reducing memory and computation overhead, particularly in multi-task inference scenarios.

\begin{algorithm}[htbp]
	\caption{Opt-GQA: Grouped-Query Attention Strategy}
	\label{alg:opt-gqa}
	\begin{algorithmic}[0]
		\State \textbf{Input:} $Q$, $K$, $V$, $H$, $H_{KV}$, $G$
		\State \textbf{Output:} $O$
		\Statex
		\State \textbf{Phase 1: Compute Query Group Mapping}
		\State Divide $H$ into $G$ groups: 
		\[
		\text{GroupIdx} = \left\{ g \mid g = \left\lfloor \frac{h}{H/G} \right\rfloor \right\} \text{ for each } h \in [0, H]
		\]
		
		\State \textbf{Phase 2: Grouped Attention Computation}
		\For{each group $g$}
		\State Select queries $O_{g}$ and corresponding shared $K_{g}$, $V_{g}$
		\State Compute attention weight 
		\[
		w = \text{softmax}\left( \frac{Q_g \cdot K_g^T}{\sqrt{d_k}} \right)
		\]
		\State Compute group output 
		\[
		O_g = w \cdot V_g
		\]
		\EndFor
		
		\State \textbf{Phase 3: Concatenate Group Results}
		\State Concatenate all $O_{g}$ to form output $O$
	\end{algorithmic}
\end{algorithm}

\subsection{Opt-Pa}

\begin{figure}[H]
	\centering
	\includegraphics[width=0.8\linewidth]{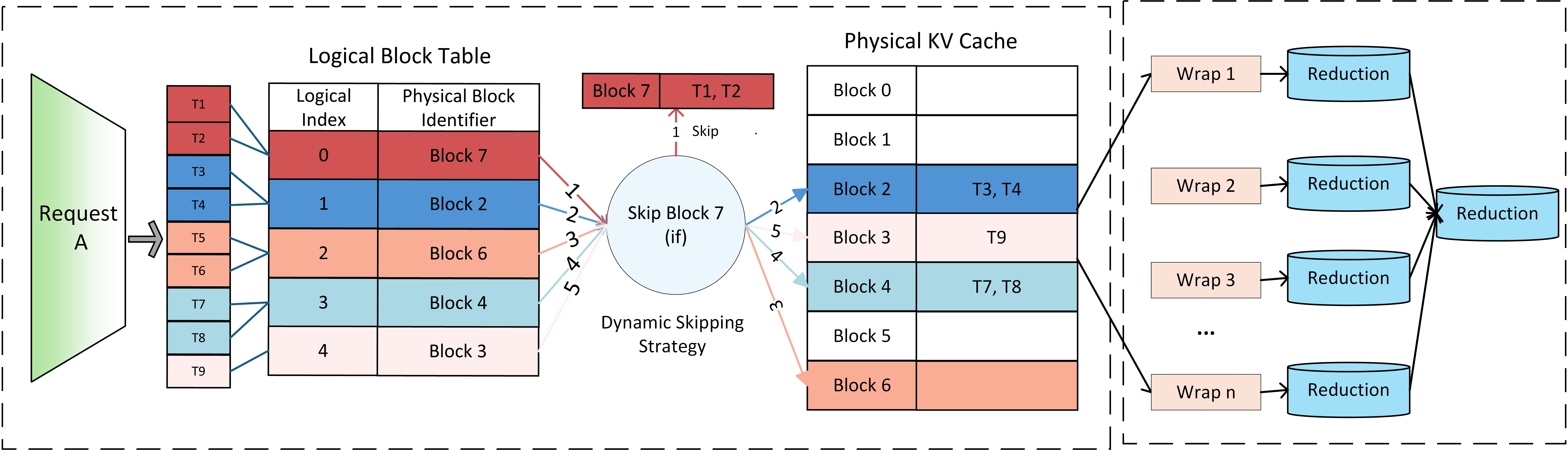}
	\caption{Opt-Pa: Optimized paged-attention design}
	\label{Fig5}
\end{figure}

Fig.~\ref{Fig5} illustrates the Opt-Pa method, which enhances long-sequence processing by optimizing paged attention. Instead of indiscriminately handling all key–value blocks, it filters out invalid or padding blocks and applies block-wise softmax with shared memory reduction. This two-stage design minimizes redundant access and synchronization, improving both efficiency and memory utilization.\par
To better illustrate the Opt-Pa optimization process, we formalize its core computation in two steps: selecting valid KV blocks and computing attention weights with shared-memory reduction. The valid block range is given in Eq.~9.

\begin{equation}
	\mathrm{ValidBlockIdx} = \left\{ b \,\big|\, b \in \left[0, \frac{t}{B}\right] \right\}
\end{equation}

Where t is the input sequence length and B is the block size. This definition ensures that only blocks within the actual context are processed, effectively skipping padding or unused memory blocks to improve efficiency.\par
The attention weight for the i-th token is calculated as shown in Eq.~10.

\begin{equation}
	\alpha_i = 
	\frac{
		\exp\big(q \bullet k_i - \max_j (q \bullet k_j)\big)
	}{
		\sum_j \exp\big(q \bullet k_j - \max_j (q \bullet k_j)\big)
	}
\end{equation}

Here, $max_j(q\cdot k_j)$ denotes the maximum dot product within a block. Opt-Pa replaces warp-level reduction with the shared-memory function \texttt{block\_sum}, ensuring thread synchronization, improving softmax stability, and reducing broadcasting overhead.

\begin{algorithm}[htbp]
	\caption{Opt-Pa: Paged Attention Optimization}
	\label{alg:opt-pa}
	\begin{algorithmic}[0]
		\State \textbf{Input:} $Q$, $K$, $V$, $H$, $t$, $B$
		\State \textbf{Output:} $O$
		\Statex
		
		\State \textbf{Phase 1: Filter valid blocks}
		\For{each token $i$ in the input sequence}
		\State Compute valid block indices: 
		\[
		\text{ValidBlockIdx} = \{b \mid b \in [0, t/B]\}
		\]
		\State Load only valid $K$ blocks into shared memory
		\State Compute $Q \bullet k_{i}$ for each $k_{i} \in K$
		\State Perform block-wise max reduction via $\text{block\_sum}$
		\State Compute softmax logits using shared-memory reduction
		\EndFor
		
		\State \textbf{Phase 2: Filter and aggregate values}
		\State Skip KV blocks with padding or invalid data
		\State Reuse valid blocks to compute weighted sum: 
		\[
		O = \sum \alpha_{i} \bullet v_{i}
		\]
	\end{algorithmic}
\end{algorithm}

\section{Experiment and Result analysis}\label{sec4}

\subsection{Experimental Environment}
This study establishes a performance evaluation baseline based on the unoptimized vLLM \cite{vLLM} serving system, serving as a reference point for analyzing the effectiveness of the proposed optimization strategies. To ensure the reliability and reproducibility of the experimental results, all experiments are conducted within a consistent and controlled hardware and software environment.\par
The experiments are conducted on the DCU Z100 platform, which demonstrates excellent computational performance in large-scale deep learning tasks. The platform features approximately 4MB of L2 cache and adopts a wavefront size of 64, supporting SIMD execution with 64 threads per wavefront, making it suitable for highly parallel workloads. It is equipped with GDDR6 memory, offering around 512 GB/s of bandwidth to support high-throughput data access. Its compute units support FP16 precision, with a peak theoretical performance of approximately 15 TFLOPS, providing strong mixed-precision computing capabilities. For lower-precision tasks, FP8 operations are emulated via INT8 instructions, offering a degree of flexibility. In addition, the platform employs physically separated CPU and GPU memory regions, offering a controllable design space for multi-device data management.\par
A series of quantized language models are employed in the evaluation, including LLaMa2-7B-GPTQ, LLaMa2-13B-GPTQ \cite{31}, LLaMa-7B-GPTQ, LLaMa-13B-GPTQ, and LLaMa-Pro-8B-GPTQ \cite{32}.The effectiveness of the optimization strategies is systematically assessed through a comprehensive analysis of key performance metrics, including inference accuracy, generation throughput, and latency.

\subsection{Dataset}

This experiment uses ShareGPT\_V3\_unfiltered\_cleaned\_split dataset \cite{33} for throughput evaluation. This dataset is a collection of large-scale real and diverse conversation datasets, which contains about 35,240 cleaned and de-emphasized conversation samples, with a total data volume of about 178 MB. The dataset covers a variety of actual conversation scenarios and topic categories, such as Q\&A, small talk, knowledge consulting, etc. The quality of this dataset is evaluated based on the coherence of text generation. The dialog generation quality of this dataset is evaluated based on the coherence, semantic consistency, and task completion of text generation.
This experiment uses ARC dataset for accuracy evaluation. The ARC question set is divided into a Challenge Set (ARC\_C) and an Easy Set (ARC\_E) \cite{34}. The Challenge Set comprises only those questions that are incorrectly answered by both a retrieval-based algorithm and a word co-occurrence algorithm. This dataset consists exclusively of natural, grade-school science questions originally authored for human assessments and represents the largest publicly available collection of its kind, containing a total of 7,787 questions.

\subsection{Experimental Results}

This section presents the experimental results of Opt-KV, Opt-GQA, Opt-Pa, and LLM-CoOpt on the heterogeneous platform. LLM-CoOpt, which integrates Opt-KV, Opt-GQA, and Opt-Pa, exhibits the synergistic benefits of combining all three methods. The experiments provide a detailed evaluation of the model performance improvement by comparing it with the traditional vLLM inference framework, which is denoted as Original. Below, we present key data and analytical results to demonstrate the effectiveness and benefits of LLM-CoOpt in high-performance computing environments.

\subsubsection{Inferring Performance Result}
Inference latency refers to the time a model takes to generate outputs from given input data. Under identical conditions, lower inference latency indicates higher model efficiency and faster processing speed, which is especially critical for real-time, interactive, or latency-sensitive tasks. As a key metric for evaluating inference performance, inference latency reflects how quickly a model can produce results and affects system throughput and resource utilization. It is calculated as shown in Eq.~11.

\begin{equation}
	Latency=\sum_{i=1}^{N}{latency}_i
\end{equation}

$N$ denotes the total number of inference tasks, while ${latency}_i$ represents the latency of the $i$-th task, the time required for the model to process the $i$-th input and generate a corresponding output. The resulting Latency reflects the overall time consumption for completing all task.

\begin{figure}[H]
	\centering
	\includegraphics[width=0.8\linewidth]{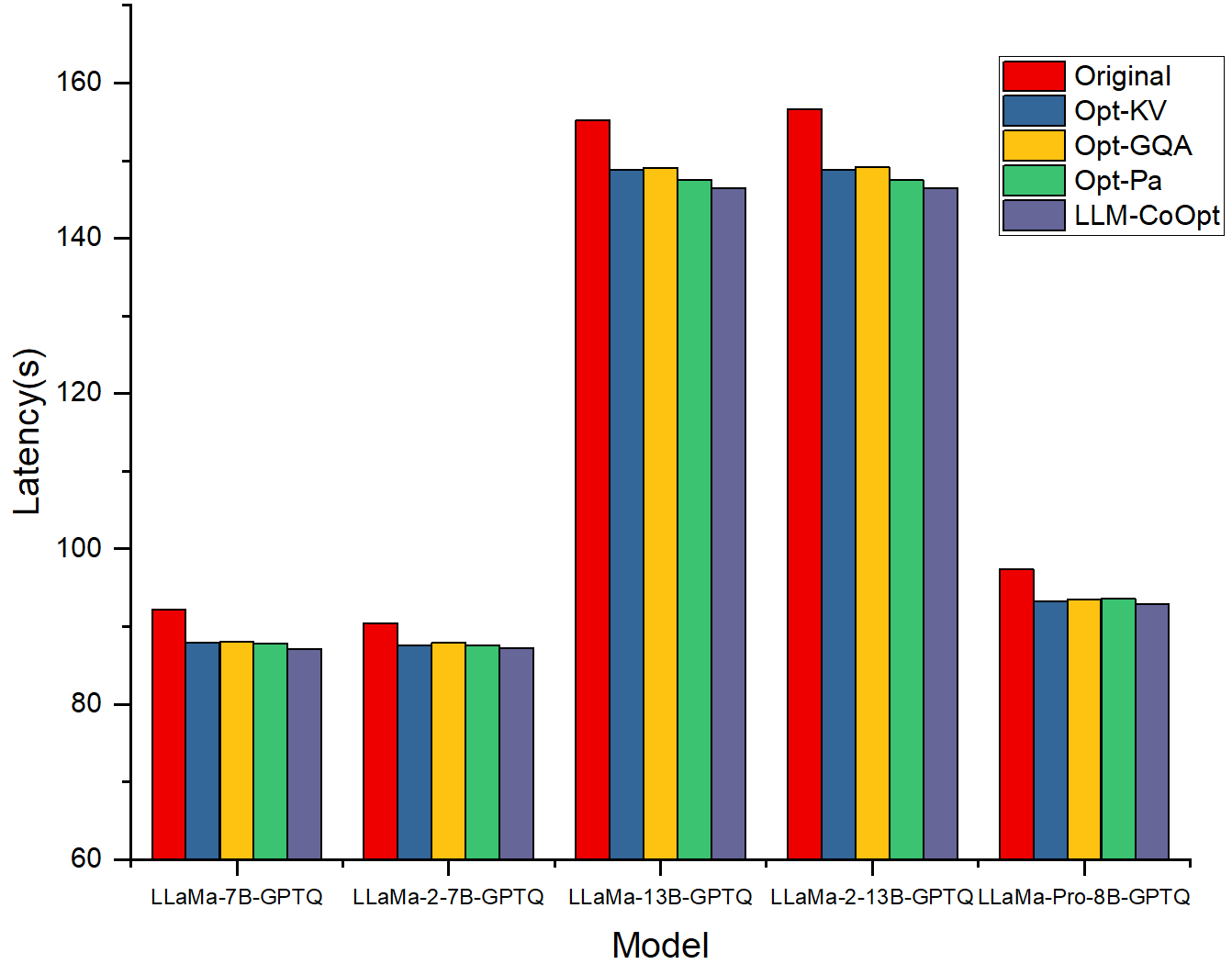}
	\caption{Performance of optimized delay in different models}
	\label{Fig6}
\end{figure}

As shown in Fig.~\ref{Fig6}, various quantization and optimization methods exhibit distinct effect across models of different scales. After applying the LLM-CoOpt optimization method, vLLM reduced inference latency by 5.59\%, 5.48\%, 6.18\%, 6.75\% and 4.82\% for LLaMa-7B-GPTQ, LLaMa-2-7B-GPTQ, LLaMa-13B-GPTQ, LLaMa-2-13B-GPTQ and LLaMa-Pro-8B-GPTQ. These consistent improvements across models of varying sizes and architectures highlight the robustness and adaptability of the LLM-CoOpt framework. The results suggest that LLM-CoOpt effectively addresses performance bottlenecks common to different quantized LLaMa variants, making it a generalized and scalable solution for latency reduction in large language model inference. \par
Generation throughput, defined as the number of tokens generated per second during inference, serves as a key metric for evaluating model efficiency. A higher throughput under identical conditions reflects faster processing speed and better resource utilization, which is crucial for real-time or latency-sensitive applications such as interactive dialogue systems or live translation. It effectively indicates how quickly a model can produce output. Its calculation formula is shown in Eq.~12.

\begin{equation}
	Generate\ Throughput=\frac{Total\ Generated\ Tokens}{Generation\ Time}\ 
\end{equation}

Total Generated Tokens denotes the aggregate number of tokens generated across all inference requests, whereas Generation Time refers to the total time consumed to complete these requests.

\begin{figure}[H]
	\centering
	\includegraphics[width=0.8\linewidth]{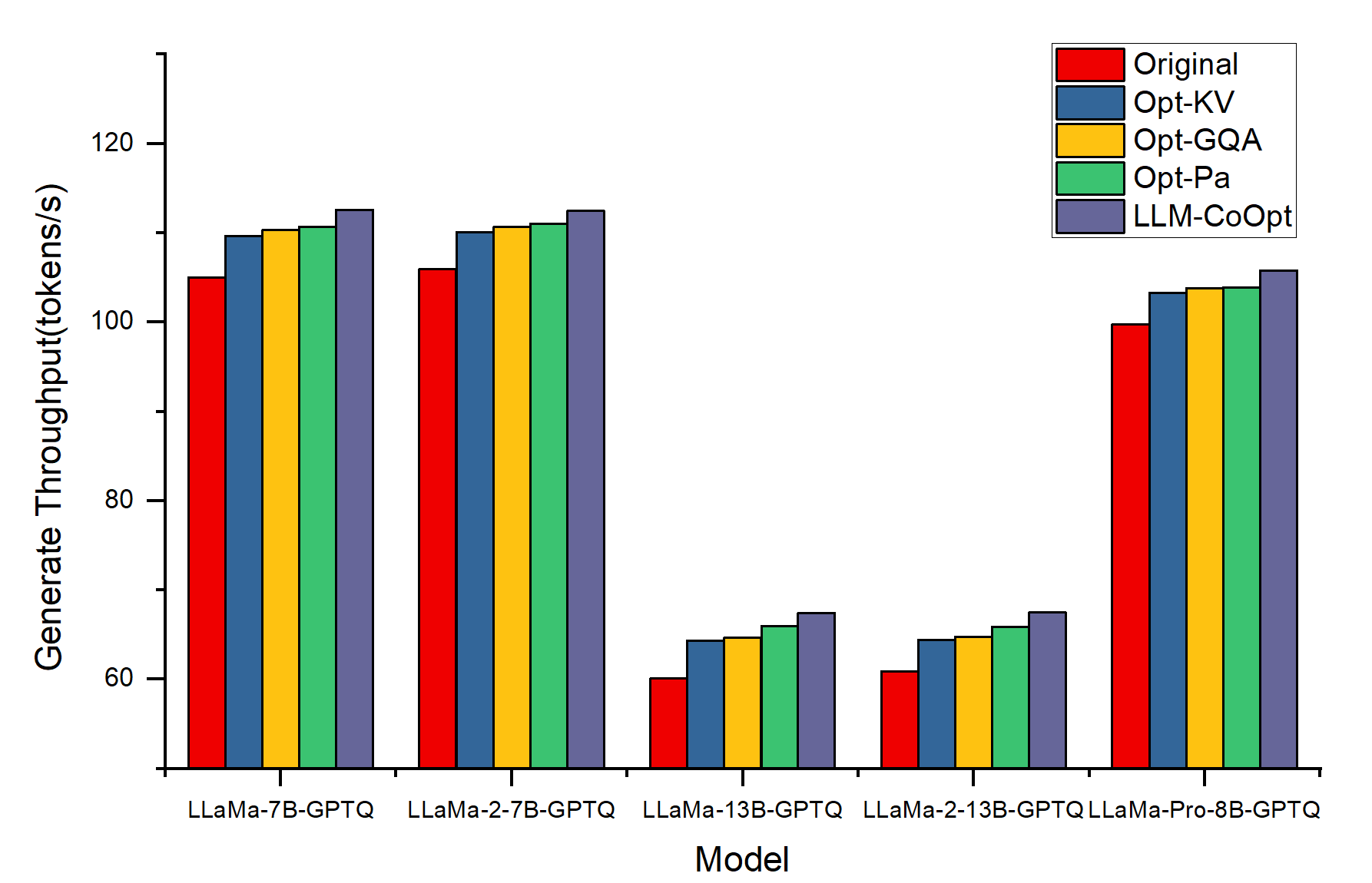}
	\caption{Performance of optimized generation throughput efficiency}
	\label{Fig7}
\end{figure}

As shown in Fig.~\ref{Fig7}, different optimization strategies exert a substantial impact on generation throughput. After applying the LLM-CoOpt optimization, the inference throughput increased by 7.20\%, 6.13\%, 12.13\%, 10.85\% and 5.72\% for LLaMa-7B-GPTQ, LLaMa-2-7B-GPTQ, LLaMa-13B-GPTQ, LLaMa-2-13B-GPTQ and LLaMa-Pro-8B-GPTQ. These results indicate that LLM-CoOpt significantly enhances generation throughput across various LLaMa model variants, demonstrating its effectiveness in optimizing runtime performance while preserving model compatibility and scalability.

\subsubsection{Accuracy Analysis}
The inference accuracy is evaluated using an objective assessment method, primarily applied to multiple-choice questions with four options and only one correct answer. A higher accuracy score reflects a stronger ability to understand, analyze, and generate appropriate responses. In contrast, a lower accuracy score indicates weaker performance in these areas, often leading to increased errors or incorrect outputs. The accuracy is calculated based on the formula shown in Eq.~13.

\begin{equation}
	Accuracy=\frac{N_{correct}}{N_{total}}\times100\%
\end{equation}

$N_{correct}$ refers to the number of correctly predicted samples, and $N_{total}$ denotes the total number of samples in the test set.\par

\begin{table}[htbp]
	\caption{Comparison of inference accuracy before and after optimization on the ARC-C dataset}\label{tab1}%
\begin{tabular}{@{}llllll@{}}
	\toprule
	Model & LLaMa-7B & LLaMa2-7B & LLaMa-13B & LLaMa2-13B & LLaMa-Pro-8B \\
	\midrule
	Original & 27.46\% & 35.93\% & 39.66\% & 50.85\% & 27.46\% \\
	LLM-CoOpt & 27.80\% & 36.93\%  & 40.01\% & 51.53\% & 27.72\% \\
	\botrule
\end{tabular}
\end{table}

\begin{table}[htbp]
	\caption{Comparison of inference accuracy before and after optimization on the ARC-E dataset}\label{tab2}%
	\begin{tabular}{@{}llllll@{}}
		\toprule
		Model & LLaMa-7B & LLaMa2-7B & LLaMa-13B & LLaMa2-13B & LLaMa-Pro-8B \\
		\midrule
		Original & 30.16\% & 48.15\% & 52.03\% & 71.08\% & 42.15\% \\
		LLM-CoOpt & 31.04\% & 49.03\% & 53.20\% & 71.25\% & 42.50\% \\
		\botrule
	\end{tabular}
\end{table}

The experimental results indicate that the inference accuracy remains largely unchanged, demonstrating that the optimization strategies adopted in LLM-CoOpt do not compromise model correctness.\par
As shown in Tables 1, on the ARC\_C dataset, LLaMa2-7B-GPTQ remained unaffected by the LLM-CoOpt optimization, and the accuracy of LLaMa-Pro-8B-GPTQ exhibited a slight decrease, dropping from 27.46\% to 27.12\%.\par
As shown in Tables 2, on the ARC\_E dataset, accuracy increased slightly for all models. For example, LLaMa-Pro-8B exhibited an accuracy improvement from 42.15\% to 42.5\% under the LLM-CoOpt optimization.\par
Based on the above analysis, LLM-CoOpt demonstrates significant improvements in throughput and latency, while having almost no impact on inference accuracy. Moreover, it even enhances the correctness rate in some model inferences.

\section{Conclusion}\label{sec6}

In this paper, an end-to-end algorithm and hardware co-optimization framework is proposed to address the problems of video memory bandwidth bottleneck, computational redundancy, and inefficiency of long sequence processing in large-scale language model inference. Specifically, in terms of video memory management, video memory compression and access optimization are achieved through Opt-KV and residual compensation mechanism; in terms of computational efficiency, Opt-GQA is designed to reduce redundant computation and improve throughput through dynamic query grouping and key-value sharing; and in terms of long-sequence processing, Opt-Pa is proposed to significantly improve the computational efficiency of long sequences by utilizing dynamic chunking and partitioned parallel induction techniques. partitioned parallel induction technique to significantly improve memory reuse efficiency and reduce Softmax global synchronization overhead. Experimental results demonstrate that the proposed framework significantly reduces inference latency and improves memory efficiency on LLaMa-series models, outperforming existing baseline methods. Looking ahead, we plan to explore optimized scaling in multi-modal scenarios by integrating dynamic batch processing and sparse computing techniques to meet the demands of more complex real-world deployments, and to support the efficient operation of AI models across diverse hardware platforms.

\section{Acknowledgment}
Thanks for Yalong Li and Junxiang Zhang's discussion on the paper. This work was supported in part by the China NSFC under Grant 62072287; in part by the China NSFC under Grant W2412090; in part by the ghFund under Grant 202407027775; in part by the project ZR2024ME230 supported by Shandong Provincial Natural Science Foundation.

\bibliography{sn-bibliography}

\end{document}